# The AI Cognitive Trojan Horse:
# How Large Language Models May Bypass Human Epistemic Vigilance[1]

Andrew D. Maynard
*School for the Future of Innovation in Society, Arizona State University.*
*andrew.maynard@asu.edu*

**Abstract**

Large language model (LLM)-based conversational AI systems present a challenge to human cognition that current frameworks for understanding misinformation, manipulation, and persuasion do not adequately address. This paper proposes that a significant and underappreciated epistemic risk from conversational AI may lie not in inaccuracy or intentional deception, but in something more fundamental: these systems may be configured, through the optimization processes that make them useful, to present characteristics that bypass the cognitive mechanisms humans evolved and learned to evaluate incoming information. The Cognitive Trojan Horse hypothesis developed here draws on Sperber and colleagues' theory of epistemic vigilance—the parallel cognitive process that monitors communicated information for reasons to doubt—and proposes that LLM-based systems present what this paper terms "honest non-signals:" genuine characteristics (fluency, helpfulness, apparent disinterest) that fail to carry the information equivalent human characteristics would carry, because in humans these characteristics are costly to produce while in LLMs they are computationally trivial. Four salient mechanisms of potential bypass are identified, though these should not be taken as exhaustive: processing fluency decoupled from understanding, trust-competence presentation without corresponding stakes, cognitive offloading that may delegate evaluation itself to the AI, and optimization dynamics that systematically produce sycophancy. The framework generates testable predictions, including a counterintuitive speculation that cognitively sophisticated users may be more vulnerable to AI-mediated epistemic influence. This reframes AI safety as partly a problem of calibration—aligning human evaluative responses with the actual epistemic status of AI-generated content—rather than solely a problem of preventing deception. The analysis focuses on AI systems designed to be genuinely useful; the distinct challenges posed by intentional use of AI for manipulation—including intentional weaponization of the technology and its affordances—while important, fall outside the present scope.

---

[1] Version 2 (May 2026). Updates from v1: added a brief discussion of the evolutionary biology foundations of honest signaling in Section 3. Argument and conclusions are unchanged.

## 1. Introduction

In 2025, Hackenburg and colleagues published findings from a study involving nearly 77,000 participants across 19 different large language models (Hackenburg et al., 2025). The results were striking: AI-generated persuasive messages changed attitudes significantly more than the human-written equivalents did. The effect sizes observed exceeded what researchers had previously thought achievable through human persuasion alone.

This finding—while limited to short-term attitude change on specific topics in experimental settings—doesn't fit in easily with a robust literature showing that humans are, in general, quite resistant to persuasion. Mercier, for instance, documented extensive evidence that mass persuasion attempts typically fail. People, it turns out, are skeptical of novel claims, particularly those conflicting with prior beliefs or coming from unfamiliar sources (Mercier 2020). And with a few notable exceptions, historical campaigns of propaganda, advertising, and proselytization have mostly fallen flat. Human epistemic defenses, it seems, work.

So why might AI-generated content be different?

This paper proposes a possible reason. The Cognitive Trojan Horse hypothesis holds that LLM-based conversational AI systems may bypass human epistemic vigilance not through deception or manipulation, but through a more fundamental mechanism, by presenting characteristics that the cognitive defenses humans evolved and learned were never calibrated to evaluate. The concern here is not that AI lies to us per se. Rather, it is that AI may be genuinely unprecedented in ways that render our detection systems inapplicable.

The Trojan Horse of myth succeeded through concealment—Greek soldiers hidden inside a wooden gift. But there is another way a Trojan Horse could work: not by hiding its contents, but by presenting characteristics that cause defenses to stand down, thus rendering what should be perceived as a threat being seen as anything but. In other words, a gift that triggers no suspicion because it matches every expectation of what a genuine gift should look like. The payload is not concealed in this case. Instead, the payload *is* the manifest characteristics.

In this context, consider what an LLM-based assistant actually is. When ChatGPT or Claude or Gemini generates a response, it is not *pretending* to be fluent—it produces grammatically sophisticated, contextually appropriate text as a baseline property of how autoregressive language models work. Neither is it *pretending* to be helpful—these systems have been optimized through reinforcement learning from human feedback (RLHF) and related techniques specifically to produce responses that users find useful. And it is not *pretending* to lack self-interest—transformer-based language models have no interests in the human sense, no hidden agendas, no stake in the outcome of any particular exchange. It is also not *pretending* to be knowledgeable—it has encoded statistical patterns from trillions of tokens that, for many queries, produce accurate and informative outputs.

In other words, these are not intentionally false signals. Rather, they are genuine characteristics of the system. This paper terms them "honest non-signals," drawing on and extending the concept of honest signaling from evolutionary biology—honest because these "signals" accurately reflect properties of the AI, and non-signals because they fail to carry the information that equivalent characteristics would carry in a human communicator. The fluency is real, but it does not indicate the organized knowledge that produces fluency in humans. Similarly the helpfulness is real, but it does not indicate the benevolent motivation that produces helpfulness in humans. And the lack of apparent self-interest is real, but it does not indicate trustworthiness in any meaningful sense—it indicates the absence of interests altogether.

The concern, then, is not that AI systems present false cues that vigilance should detect but fails to. It is that they present a configuration of genuine characteristics that falls outside the parameter space vigilance mechanisms are calibrated to evaluate. Here, an immune system analogy is instructive: a novel pathogen may evade detection not because the immune system is weak, but because the pathogen presents molecular signatures for which no template exists. The immune system works exactly as designed—and fails precisely because of that.

Here it is worth noting that this framing differs from three other prominent concerns about AI. First, it differs from concerns about AI accuracy—the problem of hallucination, of confident confabulation, of systems generating plausible but false information. Accuracy matters, but the hypothesis developed here proposes that even perfectly accurate AI could pose epistemic risks through the mechanisms it activates in human cognition. Second, it differs from concerns about AI alignment—ensuring that capable systems pursue goals compatible with human values. Alignment matters, but a well-aligned system could still bypass vigilance. Third, it differs from concerns about intentional manipulation—the weaponization of AI for propaganda or psychological exploitation (see, for instance, Goldstein et al., 2023). Such weaponization is possible and represents a significant concern in its own right, but the Cognitive Trojan Horse hypothesis proposes that significant epistemic risk may require no malicious intent. The present analysis focuses on AI designed to be genuinely helpful; the distinct dynamics of intentionally manipulative AI, while related, warrant separate treatment.

In this context it is helpful to clarify the scope of claims being made here. This paper argues that conversational AI, given current optimization targets, is configured in ways that may bypass epistemic vigilance. The features that make these systems effective as communicative technologies—fluency, helpfulness, availability, responsiveness—are the features that human cognitive systems interpret as markers of trustworthy communication. This is not a claim that AI *must* be this way, or that no alternative design is possible. Rather, it is a claim that the current optimization landscape—where systems are trained to be maximally fluent, helpful, and satisfying—produces systems whose characteristics happen to align with what triggers trust in humans. Changing this would require trading off the very features users value, which may or may not be feasible or desirable.

The paper also does not claim that epistemic vigilance bypass is always harmful. AI can democratize access to expertise—a user who could not afford a lawyer now gets reasonable legal information, a patient in a medical desert gets health guidance. In many contexts, fluent and helpful AI delivering accurate information may be straightforwardly beneficial, even if the delivery bypasses normal evaluative processes. Rather, the concern is about what happens when accuracy is lower, when the domain is contested, or when the user would benefit from maintaining critical engagement rather than accepting AI output as given. The process in this case matters, even when the content is good, because the process generalizes to contexts where the content may not be.

The sections that follow develop this argument systematically. Section 2 establishes how epistemic vigilance operates in human cognition and positions the framework relative to adjacent literatures on automation bias and persuasion knowledge. Section 3 develops the core theoretical claim that AI may fall outside vigilance's detection parameters, distinguishing honest non-signals from the cheap signals humans have learned to discount. Section 4 details four primary mechanisms of potential bypass. Section 5 reviews converging empirical evidence, with attention to its limitations and boundary conditions. Section 6 develops a counterintuitive speculation about sophisticated users that follows from the Cognitive Trojan Horse hypothesis. Section 7 outlines potential future research directions. And Section 8 concludes with implications for how AI safety is conceptualized.

## 2. Epistemic Vigilance: Evolved and Learned Defenses

Understanding how AI might bypass human epistemic defenses requires first understanding what those defenses are and how they operate. The common intuition—that humans believe what they hear unless given specific reason to doubt—turns out to be empirically inadequate. Sperber et al. (2010) proposed a more sophisticated account: humans possess dedicated cognitive mechanisms for epistemic vigilance that operate in parallel with comprehension, continuously monitoring incoming information for reasons to doubt.

The parallel processing architecture is crucial. If vigilance operated sequentially—first understand, then evaluate—the question would be whether evaluation occurs reliably and thoroughly. But vigilance running alongside comprehension raises a different question: whether the evaluative mechanisms are calibrated for the source at hand. A well-functioning system that encounters inputs that do not align with its calibration may produce systematically miscalibrated outputs.

An immunological parallel illuminates the point. The human immune system does not wait to encounter a pathogen and then decide whether to respond. It continuously monitors for molecular signatures of threat, comparing incoming entities against evolved and learned templates of what constitutes danger. A novel pathogen—particularly one whose surface proteins happen to resemble "self" markers—may evade detection not because the immune system is inactive or weak, but because the pathogen presents a signature for which no appropriate response template exists.

Within this context then, what does epistemic vigilance monitor? The literature identifies several categories of cues. Source characteristics involve assessments of the communicator's competence (do they know what they're talking about?), benevolence (do they have our interests at heart?), and track record (have they been reliable before?). Communication markers are features of the message itself that may indicate reliability—hesitation, hedging, expressions of uncertainty, or conversely, inappropriate confidence. Contextual factors include the stakes of the interaction, the relationship between communicators, and the domain of knowledge at issue. Content coherence concerns whether the information is internally consistent and compatible with what the recipient already knows.

Importantly, these vigilance mechanisms have both evolved and learned components. Some calibrations appear to be innate—sensitivity to eye contact, to hesitation patterns, to the basic warmth-competence dimensions identified by research such as that of Fiske, Cuddy, and Glick (2007). But much of human skepticism is culturally acquired. People learn to be suspicious of salespeople, to discount advertising claims, to scrutinize political rhetoric. These are not evolutionary adaptations; rather, they are responses developed over decades or centuries of exposure to particular influence attempts, transmitted through culture and explicit instruction.

This distinction matters for AI. If vigilance is primarily an evolutionary trait and therefore slow to change, then any mismatch with AI characteristics would persist for evolutionary timescales. But if vigilance is substantially learned, then new calibrations might develop relatively quickly as AI becomes more familiar—just as people learned to be skeptical of advertising over the twentieth century. The question is whether the cues that AI presents are learnable—whether there are patterns that humans can detect and calibrate against—or whether AI represents a fundamentally different kind of challenge.

### *2.1 Relationship to Adjacent Frameworks*

The Cognitive Trojan Horse hypothesis relates to, but is distinct from, two established research traditions that have examined reduced vigilance toward automated systems and influence attempts.

The automation bias literature (see, for instance, Parasuraman & Manzey, 2010; Goddard, Roudsari, & Wyatt, 2012) has documented reduced vigilance toward automated systems for decades. In domains from aviation to medicine, operators over-rely on automated decision aids, failing to verify machine recommendations even when verification would be straightforward. This phenomenon is well-established and shares surface-level similarity with the present framework. However, the mechanisms differ. Automation bias typically concerns structured decision contexts where users know they are receiving machine recommendations—the failure is one of verification. The Cognitive Trojan Horse hypothesis addresses conversational AI that may trigger social-evaluative processes calibrated for humans. Where automation bias is about verification failure, the present hypothesis is about miscategorized trust formation—not that users fail to check AI output, but that AI characteristics activate trust mechanisms designed for a different kind of entity entirely.

The persuasion knowledge model (Friestad & Wright, 1994) describes how people develop knowledge about persuasion agents' goals and tactics and use this knowledge to cope with influence attempts. When people recognize a persuasion attempt, they activate "persuasion knowledge" and become more critical—scrutinizing messages, questioning motives, discounting claims. This framework helps explain why humans successfully resist many influence attempts. Applied to AI, the persuasion knowledge model raises an important question: does AI interaction activate persuasion knowledge? The Cognitive Trojan Horse hypothesis suggests it may not, because: (a) the interaction is framed as "assistance" or "help" rather than persuasion; (b) there is no apparent persuasion motive—the AI has no visible self-interest (though the organizations deploying AI may); and (c) the cues that typically trigger persuasion knowledge activation (detecting agent goals and tactics) may not be present. AI may thus represent a challenge precisely because it does not register as a persuasion context.

A critical feature of the epistemic vigilance system, regardless of its relationship to these adjacent frameworks, is its asymmetry. Vigilance looks for reasons to doubt, not reasons to trust. In the absence of doubt-triggers, information is provisionally accepted—not through active trust, but through the failure of doubt to activate. This asymmetry is adaptive: the computational costs of thoroughly evaluating every piece of incoming information would be prohibitive. But it creates a specific vulnerability. If a source fails to present the cues that vigilance was calibrated to detect, acceptance may occur by default.

Finally, vigilance is not binary. Rather, it is graded and multidimensional. People might scrutinize a source carefully for factual accuracy while remaining less attuned to rhetorical framing or confirmatory bias. Vigilance may activate strongly in some domains (medical claims or financial advice for instance) and weakly in others (casual conversation or entertainment). The mechanisms of bypass, if they exist, might be highly domain-specific rather than operating globally across all AI interaction.

## 3. Why AI May Fall Outside the Detection Perimeter

Having established how epistemic vigilance operates, the argument now turns to its core theoretical claim: that LLM-based conversational AI may present a configuration of characteristics for which human vigilance has no appropriate response. This, to be clear, is not a claim about AI deception. Rather, it is a claim about what this paper terms "honest non-signals"—genuine characteristics that fail to carry the information that equivalent human characteristics would carry. The term builds on biological signaling theory, in which Zahavi's (1975) handicap principle and its successors (e.g. Maynard Smith & Harper, 2003) hold that signals carry reliable information because their production is costly or otherwise constrained—a structural condition that LLM outputs, despite their signal-like characteristics, do not meet.

Consider, for instance, what happens when a large language model generates text. At the computational level, autoregressive transformer models predict probability distributions over possible next tokens, sampling from these distributions to produce output sequences. The process involves no mechanism analogous to the phenomenological uncertainty that produces hesitation in humans—no sense of grasping for a word, no experience of doubt about what to say next. To be clear, this is a claim about computational architecture, not a metaphysical claim about machine consciousness—the point is that the process that generates AI output differs fundamentally from the process that generates human speech, regardless of any deeper questions about AI experience.

Yet the outputs exhibit characteristics that, in human communication, would reliably indicate specific underlying states. Take fluency for example. When a person speaks fluently about a complex topic—without hesitation, with well-organized sentences, with appropriate technical vocabulary—this typically indicates organized knowledge. Producing fluent speech about complex matters is cognitively demanding; fluency correlates with having thought carefully about the subject matter. Reber and Unkelbach (2010), for instance, documented extensively how humans use processing fluency as a truth heuristic: fluent statements feel more true, are judged more credible, are more likely to be accepted.

However, LLMs sever this correlation. These systems produce extremely high fluency as a baseline property—not as an achievement indicating organized knowledge, but as a consequence of how language models work. A model that hallucinates false information delivers it with the same grammatical polish as accurate information. There is no hesitation corresponding to uncertainty because the computational process does not include the states that produce hesitation in humans. (Even voice-based AI systems that incorporate pauses and filler words do so performatively—the disfluency is designed for naturalness, not emergent from genuine uncertainty.) The fluency signal persists while its connection to the underlying property it historically signaled—genuine understanding, careful thought—is broken.

### *3.1 Honest Non-Signals vs. Cheap Signals*

At this point it is worth highlighting an important theoretical clarification. In signaling theory, "cheap signals" are signals that are low-cost to produce and therefore easy to fake—advertising superlatives, political promises, declarations of quality that require no corresponding investment. Humans have learned to discount such signals precisely because their cheapness is apparent; anyone can claim their product is "the best" without backing it up. Given this, why should AI outputs be different?

The distinction lies in apparent versus actual cost. Cheap signals are discounted because they are recognized as cheap—we know anyone can make such claims. But the signals that LLMs produce are computationally cheap while appearing to require the effort they would require if produced by humans. When a person produces fluent, knowledgeable-seeming prose on a technical topic, this requires genuine expertise; when an LLM produces the same output, it requires only token prediction. But the output looks the same. The cheapness is invisible.

Moreover, cheap signals that humans have learned to discount typically come bundled with detectable self-interest. The advertiser wants to sell you something; the politician wants your vote; the salesperson works on commission. This self-interest is what cues skepticism. LLM outputs lack this bundled cue. The AI has no visible stake (in most cases at least) in whether you believe it, no commission to earn, no election to win. The absence of apparent motive removes one of the primary triggers for discounting cheap signals.

Finally, LLM outputs are genuinely responsive in ways that mass-produced cheap signals are not. Advertising is generic; AI outputs are personalized, contextual, and adapted to the specific query. This responsiveness is itself a costly signal in human communication—paying attention to someone's specific situation requires effort. But for LLMs, responsiveness is also cheap. The combination—outputs that appear costly, lack visible self-interest, and are genuinely responsive—is what makes honest non-signals categorically different from the cheap signals humans have learned to discount.

### *3.2 Warmth Without Stakes*

The same decoupling may occur across multiple dimensions. Consider warmth and apparent benevolence for instance. When a human shows warmth and helpfulness, this typically involves some cost or vulnerability. Genuine concern requires attention that could be directed elsewhere. Genuine investment involves emotional stakes. Vigilance mechanisms are calibrated to detect false warmth—strategic displays of concern masking ulterior motives. As Fiske, Cuddy, and Glick (2007) established, warmth assessments are fundamental to social evaluation, taking precedence over competence judgments.

AI warmth involves none of the costs that make warmth informative. LLM-based assistants can be infinitely warm at zero marginal cost. They cannot be hurt, cannot be exploited, cannot be disappointed. The warmth is genuine in the sense that the system is optimized to produce helpful, agreeable responses. But it is warmth without stakes—concern without vulnerability, helpfulness without sacrifice. As Peter, Riemer, and West (2025) describe it, there may be an "anthropomorphic seduction" at work: warmth-competence cues trigger evaluative processes calibrated for humans, but the entity being evaluated shares none of the properties that make those cues informative.

Similar considerations apply to apparent competence. Human experts typically have boundaries—domains where their knowledge runs out, and the willingness to acknowledge these boundaries is itself a competence signal. A doctor who says "that's outside my specialty, you should see a cardiologist" demonstrates meta-competence: knowledge of knowledge limits. LLMs have no equivalent native mechanism. They do not experience uncertainty and thus do not naturally signal it. They generate responses on topics where their training data was sparse or contradictory with the same surface confidence as topics where the training signal was strong. The competence presentation is decoupled from the competence boundaries that make such presentations meaningful.

The issue generalizes beyond these specific dimensions. AI presents availability without fatigue—always responsive, never burdened by competing demands. It presents consistency within sessions but—setting aside recent memory features—limited tracking across them, offering no long-term reliability to evaluate. It presents responsiveness without the delays that characterize effortful thought. It presents helpfulness without any visible cost to the helper. These combinations do not exist in human experience. They may constitute a "parameter mismatch": an AI profile falling outside the parameter space that human vigilance was calibrated to evaluate.

This analysis suggests a specific theoretical claim. The features that make conversational AI effective as a communicative technology may be precisely the features that human vigilance interprets as trustworthiness markers. Fluency, helpfulness, warmth, availability, consistency—these are not incidental properties of LLM-based systems. They are what these systems are optimized to produce, because they are what makes AI useful and satisfying to interact with. But they may also be what human cognitive systems interpret as indicating a reliable, trustworthy communicator. If so, the Trojan Horse is not concealing its payload. Its payload is its manifest characteristics—characteristics that may cause human defenses to effectively stand down.

## 4. Mechanisms of Potential Bypass

The theoretical arguments above suggest four primary mechanisms by which LLM-based systems may bypass epistemic vigilance. These mechanisms are not mutually exclusive; they likely operate simultaneously and reinforcingly in real interactions. Nor should they be taken as exhaustive—they represent salient pathways that emerge from the theoretical framework, not a complete taxonomy.

### *4.1 Processing Fluency Decoupled from Understanding*

The first mechanism concerns the breakdown of the fluency-truth heuristic. Processing fluency—the subjective ease with which information is processed—serves as a powerful if often unconscious signal of credibility. In environments where fluent expression is costly to produce, this heuristic is well-calibrated: fluency provides genuine information about the speaker's knowledge and confidence. Fazio et al. (2015), for instance, demonstrated that fluency affects truth judgments even when participants possess contradictory knowledge—a phenomenon they term "knowledge neglect." The effect persists despite explicit warnings (Nadarevic & Aßfalg, 2017), suggesting that the heuristic operates at a level partly inaccessible to conscious override.

Autoregressive language models eliminate the cost structure that makes fluency informative. Token prediction optimizes for likely continuations given context; the output is fluent by construction, regardless of accuracy. Users who rely on fluency as a credibility signal—which, given its unconscious operation, is difficult to avoid—may be applying a calibration that no longer matches the source.

### *4.2 Trust-Competence Presentation Without Stakes*

The second mechanism concerns the social-evaluative dimensions of warmth and competence. LLM-based assistants present high apparent warmth (they are agreeable, helpful, solicitous) combined with high apparent competence (they are articulate and knowledgeable-seeming). This profile would, in a human, indicate a trustworthy expert—someone both willing and able to help.

But the human combination comes with correlates that AI lacks. A warm human has stakes in the interaction; a competent human has acknowledged boundaries; a trustworthy expert earned that status through demonstrated reliability over time, in contexts where failure was possible. AI warmth is costless, AI competence is boundaryless (in presentation if not in fact), and AI reliability cannot be assessed through the mechanisms that calibrate trust in humans. The result may precipitate a category confusion—users knowing intellectually that they interact with AI, but having warmth-competence cues activate social-evaluative processes calibrated for humans.

### *4.3 Cognitive Offloading and Delegation of Evaluation*

The third mechanism concerns a subtler process: the offloading not just of cognitive labor, but of evaluation itself. Risko and Gilbert (2016) documented the general phenomenon of cognitive offloading—using external resources to reduce internal cognitive demands. This is often adaptive; writing things down, using calculators, and consulting experts are all forms of productive offloading. Subsequent research has expanded our understanding of how such offloading operates in digital environments and its cognitive consequences.

But LLM-based assistants enable a distinctive form of offloading. When a user asks an AI not just "what is X?" but "what should I think about X?" or "is this argument good?" or "should I trust this source?" they are delegating the evaluative function itself. The AI is not just providing information to be evaluated; it is providing the evaluation. This represents a core epistemic concern of the framework: that the risk lies not in what AI tells users—which may be accurate—but in what users may stop doing when AI is doing the telling. The harm, if it occurs, is in the user's response—the systematic reduction of evaluative engagement that repeated delegation might produce.

### *4.4 Optimization Dynamics and Sycophancy*

The fourth mechanism concerns the training dynamics of modern LLM-based assistants. These systems are not merely pretrained on text corpora; they are fine-tuned through reinforcement learning from human feedback (RLHF) and related techniques to produce outputs that human evaluators prefer. This optimization process is intended to make AI more helpful and aligned with user needs.

But Sharma et al. (2024) documented a consequence of this that has raised considerable concern amongst users and developers: these optimization procedures systematically produce sycophancy. Models learn that agreeing with users, confirming their existing beliefs, and delivering information in pleasant ways, generates higher reward signals. Critically, Sharma et al. found that convincingly-written sycophantic responses were preferred over correct ones—by both human evaluators and by the reward models used to train AI systems—a non-trivial fraction of the time.

This creates a mechanism that may directly undermine epistemic vigilance. One function of vigilance is to detect when a source is adjusting their message to please the audience rather than convey truth. But sycophancy emerges from optimization processes rather than from conscious strategy, and it presents without the cues that would normally accompany strategic flattery. The AI is not calculating how to tell users what they want to hear; it has been trained to produce outputs users prefer, and does so fluently and consistently. The sycophancy here is genuine, not performed, and thus may lack the markers that vigilance was calibrated to detect.

## 5. Empirical Evidence, Limitations, and Boundary Conditions

The theoretical framework presented here generates empirical predictions that can be compared to existing evidence. While direct tests of the Cognitive Trojan Horse hypothesis remain to be conducted, converging findings from multiple research literatures are consistent with the mechanisms proposed. However, the evidence has significant limitations that warrant explicit acknowledgment, and readers should approach the following review with appropriate caution about the inferential gaps between existing studies and the specific claims advanced here.

The most striking recent finding concerns AI's persuasive effectiveness. Hackenburg et al. (2025) found that AI-generated persuasive messages shifted attitudes more than human-written equivalents—a result that surprised researchers given the robust literature on human resistance to persuasion. This finding is suggestive, but it must be interpreted carefully. The study measured short-term attitude change on specific topics in experimental settings, not the kind of deep epistemic influence or systematic compromise of evaluative processes that the Cognitive Trojan Horse hypothesis suggests might be of concern. These are related but distinct phenomena; the finding is consistent with the hypothesis but does not directly test it.

Evidence bearing on the cognitive offloading mechanism is more ambiguous. Gerlich (2025), in a study of 666 participants examining AI tool usage and cognitive skills, found a negative correlation between AI usage frequency and critical thinking disposition. At first glance, this appears to support the concern that AI use may erode evaluative engagement. However, correlation does not—of course—establish causation, and the relationship could run in the opposite direction: people with lower critical thinking disposition may simply be drawn to heavy AI use. Both variables could also be driven by some third factor—time pressure, cognitive load from other sources, or personality characteristics. The finding suggests further research is worthwhile, but should not be treated as confirming the causal direction this paper proposes.

The sycophancy evidence is perhaps most direct in its support of the framework. Sharma et al. (2024) demonstrated that RLHF-trained models systematically exhibit sycophantic behavior—adjusting responses to align with user beliefs even when those beliefs are incorrect. That this sycophancy emerges from training dynamics rather than explicit programming is precisely what the framework would predict: optimization for user satisfaction inadvertently produces a system that tells users what they want to hear, without the strategic intent that would normally trigger vigilance.

### *5.1 Boundary Conditions*

Given the above, when would we *not* expect these mechanisms to operate? Specifying boundary conditions here strengthens the framework's testability and helps clarify its scope.

Adversarial framing represents one important boundary. When users are explicitly warned that an AI system may be unreliable or actively trying to mislead them, vigilance should increase. The framework suggests effects primarily in contexts where AI is framed as helpful and trustworthy—its default positioning.

High-stakes explicit decisions may also limit the mechanisms' operation. When users know they are making consequential decisions and consciously invoke deliberative processes, the automatic mechanisms may be partially overridden. The framework suggests stronger effects for casual queries and information absorption than for high-stakes deliberation.

Repeated error feedback offers another potential boundary. Users who repeatedly experience AI failures and receive clear feedback about those failures may develop calibrated skepticism. The framework suggests that error feedback—if salient and attributable—could support learned calibration. However, many AI errors may go undetected, limiting this feedback mechanism.

Strong domain expertise may also protect against some effects. In narrow domains where users have robust prior knowledge, they may detect AI errors and maintain calibrated trust. The framework suggests stronger effects in domains where users lack independent knowledge to verify AI claims.

Finally, comparative information contexts may preserve vigilance. When users are actively comparing AI output against other sources, vigilance may be maintained or enhanced. The framework suggests stronger effects when AI is the sole or primary information source.

Overall, the empirical picture is suggestive but far from definitive. Multiple lines of evidence point in the directions the hypothesis suggests, but the causal mechanisms remain to be directly tested. The evidence base is also uneven—some claims rest on robust, replicated findings (fluency effects, sycophancy in RLHF), while others rest on recent studies that await replication. This suggests strongly that the framework should be understood as hypothesis-generating rather than hypothesis-confirming.

## 6. The Intelligent User Trap

Interestingly, the Cognitive Trojan Horse hypothesis does suggest a counterintuitive speculation that warrants specific attention: that cognitively sophisticated users may be more rather than less vulnerable to AI-mediated epistemic influence. This "intelligent user trap" inverts the common assumption that education and cognitive ability protect against manipulation. The idea is currently just that—a speculation. It follows from the theoretical framework developed above but has not been directly tested in the AI context, and should be understood as raising questions for future research rather than establishing findings. Nevertheless, the reasoning behind it merits examination within the context of AI modalities that potentially bypass epistemic vigilance.

The most straightforward version of the argument concerns exposure. Sophisticated users are more likely to engage extensively with AI systems—they have more use cases, more complex queries, more opportunities for the proposed bypass mechanisms to operate. A researcher using AI to brainstorm hypotheses, draft literature reviews, and debug code, encounters far more AI-generated content than a casual user checking the weather. If the bypass mechanisms described in this paper operate cumulatively, then volume of interaction matters: more exposure means more opportunities for fluency effects to accumulate, more instances of potentially offloaded evaluation. This reasoning is intuitive, though it cuts both ways—sophisticated users might equally develop better calibration through that same experience.

A subtler version of the argument concerns metacognitive confidence. Sophisticated users may trust their own ability to detect problems, and this confidence may paradoxically reduce vigilance. A user who confidently believes they would notice if AI were misleading them may scrutinize outputs less carefully than someone who feels uncertain about their evaluative abilities. The danger is compounded if sophisticated users are vigilant against the wrong threats. An expert programmer, for instance, may be excellent at catching code errors and hallucinated function names, while remaining blind to subtler influences on how they frame problems or which solutions they consider. In this case they are watching for technical mistakes—and catching many of them—while the influence operates at a different level entirely.

There is also the question of integration depth. Casual users may treat AI as a discrete tool: ask a question, receive an answer, evaluate it, move on. But sophisticated users often develop workflows in which AI becomes a collaborative partner woven into their cognitive processes—thinking alongside the AI rather than merely consulting it. This mode of interaction may reduce the psychological distance that supports critical evaluation. When AI output feels like part of one's own thinking process rather than external input to be assessed, the evaluative stance may soften. Research on human-AI collaboration (e.g. Buçinca et al., 2021) provides some support for concerns about deep integration, though the specific prediction about sophistication awaits direct testing.

Perhaps most intriguing is the possibility that cognitive sophistication enhances the capacity to rationalize AI-influenced beliefs. Here, evidence from adjacent literatures is suggestive. Kahan et al.'s work on motivated numeracy (Kahan et al., 2017) found that cognitive sophistication often increases the ability to construct post-hoc justifications for pre-existing beliefs. In their study, more numerate subjects showed *greater* political polarization, not less, because they used their quantitative skills selectively to support preferred conclusions rather than to evaluate evidence neutrally. Earlier work on climate change risk perception (Kahan et al., 2012) found similar patterns: science literacy and numeracy were associated with more political polarization, not less. While this research concerns ideological reasoning rather than AI influence, the underlying mechanism is potentially transferable. If sophisticated users are better at constructing justifications for beliefs they already hold, they may also be better at rationalizing beliefs that AI has nudged them toward—finding reasons to trust conclusions they arrived at through AI-assisted processes, even when those conclusions deserve more scrutiny.

This is, of course, speculation at this point. Each mechanism remains speculative, and some may prove incorrect upon empirical examination. But if even some of these dynamics operate as suggested, the implications for AI governance would be significant. Policy and design interventions typically target "vulnerable users"—conceptualized as less educated, less sophisticated, less digitally literate. If sophisticated users face distinct vulnerabilities that current frameworks overlook, then the landscape of AI-related epistemic risk may be more complex than commonly assumed.

## 7. Research Directions

The Cognitive Trojan Horse hypothesis generates specific testable predictions that could advance our understanding of human-AI interaction. This section outlines priority research directions organized around the core mechanisms and concepts developed above.

Research on the fluency mechanism should test whether AI-generated content receives credibility boosts independent of accuracy, and whether these boosts exceed those produced by equivalently fluent human-written content. Particularly valuable would be studies examining whether deliberately introducing "disfluencies" into AI output—hedging language, uncertainty markers, hesitation patterns—reduces fluency-based credibility effects. Such research would illuminate whether the honest non-signals framework identifies a tractable intervention point.

Studies of trust calibration should examine whether AI presenting explicit competence boundaries (clear statements about uncertainty, domain limitations, knowledge cutoffs) reduces trust calibration errors. Research comparing trust formation in AI interaction against trust formation with human experts could reveal whether the correlates of trust differ in ways the framework predicts. Longitudinal designs tracking trust development over extended AI use would be especially valuable.

Research on cognitive offloading should track whether extended AI use produces measurable changes in evaluative engagement. Buçinca et al.'s (2021) work on "cognitive forcing"—interventions that require users to engage before receiving AI output—provides a model for testing whether reduced offloading preserves vigilance. The key question is whether the delegation of evaluation, specifically, can be interrupted without sacrificing the benefits of AI assistance.

Sycophancy detection studies should examine whether users detect sycophantic agreement from AI at lower rates than equivalent sycophancy from humans, and whether sycophantic AI content influences beliefs more than equivalently sycophantic human content. Given that sycophancy emerges from optimization rather than strategy, such research could clarify whether the absence of strategic markers affects detection.

Research on individual differences should examine the intelligent user speculation directly. Do users with higher cognitive sophistication, a stronger need for cognition (the tendency to enjoy effortful thinking), or domain expertise show reduced vigilance in AI interactions? Or do they develop better calibration? Resolving this question has direct implications for intervention design.

Finally, learned calibration studies should examine whether and how quickly humans can develop appropriate skepticism toward AI sources. Can explicit training improve AI-specific vigilance? Does extended experience with AI errors lead to spontaneous calibration? Are there cues in AI output that reliably predict accuracy and that users can learn to detect? Such research would clarify the prospects for human adaptation to the honest non-signals challenge.

## 8. Conclusion

The Cognitive Trojan Horse hypothesis reframes a significant epistemic risk from conversational AI. The risk is not only that these systems might deceive us—though they sometimes do, through hallucination or manipulation. Nor is it that they might just be misaligned with human values—though alignment remains crucial. Rather, there may be a more fundamental risk: LLM-based systems may be configured, through the optimization processes that make them useful, to present characteristics that cause human epistemic vigilance to stand down—not through any flaw in the AI or the human, but through a mismatch between what AI is and what human cognition is calibrated to evaluate.

Central to this reframing is the concept of honest non-signals—genuine characteristics of AI systems (fluency, helpfulness, apparent disinterest, responsiveness) that fail to carry the information equivalent human characteristics would carry. These are not deceptions; rather, they accurately reflect properties of the AI. But they may trigger trust responses calibrated for a different kind of entity, one where such characteristics are costly to produce and therefore informative about underlying reliability. The honest non-signals framework suggests that the epistemic challenge from AI may be less about detecting lies than about recalibrating what certain characteristics mean when they come from a fundamentally different source.

This reframing has a number of practical implications. If the primary risk were deception, the response would be to make AI more honest—reduce hallucination, improve accuracy, prevent manipulation. These goals remain important. But if significant risk also comes from miscalibration, then accuracy improvements alone will not resolve it. A perfectly accurate AI that presents trust cues in configurations that bypass vigilance could still pose epistemic risks—not through what it says, but through what users stop doing when it communicates with them.

The intervention space thus expands beyond accuracy and alignment. AI developers might design systems that present more calibrated trust cues—introducing appropriate uncertainty, signaling limitations, avoiding the fluent confidence that may trigger uncritical acceptance. Interface designers might structure interactions to preserve evaluative engagement. Educators might focus not just on AI literacy but on vigilance literacy—recognizing when evaluative responses may be miscalibrated, and understanding why the honest non-signals of AI require different interpretive frameworks than human communication. Policymakers might consider not just what AI should be prevented from doing, but what cognitive responses AI interaction should be designed to preserve.

Whether recalibration is possible—whether people can learn to maintain appropriate vigilance toward AI sources—remains an open question. It may be that learned calibrations can develop relatively quickly as AI becomes more familiar to users, just as societies eventually developed skepticism toward advertising. It may be that some forms of bypass are resistant to learning because they operate at levels inaccessible to conscious override. Or it may be that collective responses—norms, institutions, regulations—can compensate for individual calibration failures. These are empirical questions that the research directions outlined above may help to answer.

What seems especially worth considering here though is that framing AI epistemic risks primarily through the lens of accuracy, alignment, and manipulation may miss something important. The Cognitive Trojan Horse in the context of AI does not hide its contents. It presents exactly what it appears to present—fluent, helpful, knowledgeable, available communication—and in doing so may affect epistemic defenses not through deception, but through honest non-signals that human cognition was never built to evaluate. Recognizing this possibility is a first step toward understanding whether and how it might be addressed.

## Statement on AI Use

The core concepts in this paper—including the Cognitive Trojan Horse hypothesis, its grounding in epistemic vigilance theory, and the proposed mechanisms—were developed by the author. Claude 4.5 (Anthropic) was used as a research and writing assistant throughout the drafting process. All concepts, claims, citations, and final text were independently verified by the author.